\documentclass[12pt]{article}
\usepackage{latexsym}
\usepackage{epsf}

\newcommand{\andvol}[3]{{\bf #1}~(#3)~#2}

\newcommand{\PRL}[3]{Phys.~Rev.~Lett.~\andvol{#1}{#2}{#3}}
\newcommand{\PRD}[3]{Phys.~Rev.~\andvol{D#1}{#2}{#3}}
\newcommand{\PR}[3]{Phys.~Rev.~\andvol{#1}{#2}{#3}}
\newcommand{\NPB}[3]{Nucl.~Phys.~\andvol{B#1}{#2}{#3}}
\newcommand{\PLB}[3]{Phys.~Lett.~\andvol{B#1}{#2}{#3}}

\newcommand{\ZPC}[3]{Z.~Phys.~\andvol{C#1}{#2}{#3}}

\newcommand{\hepph}[1]{\ hep-ph/#1}

\newcommand{\etal}{\textit{et al.}}
\newcommand{\dfrac}[2]{\frac{\displaystyle #1}{\displaystyle #2}}
\newcommand{\vev}[1]{\left\langle #1 \right\rangle}
\newcommand{\dgl}{\delta g_L}
\newcommand{\dgr}{\delta g_R}
\newcommand{\ds}{\delta s^2}
\newcommand{\da}{\delta\alpha_s}
\newcommand{\GamZ}{\Gamma_Z}
\newcommand{\Gamhad}{\Gamma_\mathrm{had}}
\newcommand{\Game}{\Gamma_{e^+e^-}}

\newcommand{\sighad}{\sigma_\mathrm{had}^0}
\newcommand{\AFB}{A_\mathrm{FB}^0}
\newcommand{\tpshift}{\Delta}
\begin{document}

\renewcommand{\thefootnote}{\fnsymbol{footnote}}

\begin{titlepage}

\begin{flushright}
VPI--IPPAP--98--7\\
hep--ph/9812377\\
December 1998
\end{flushright}

\bigskip
\bigskip
\bigskip
\bigskip

\begin{center}

\textbf{\large
Constraints on Topcolor Assisted Technicolor Models
from Vertex Corrections
}

\bigskip
\bigskip
\bigskip

\textsc{Will~LOINAZ}\footnote{%
electronic address: loinaz@alumni.princeton.edu}
and
\textsc{Tatsu~TAKEUCHI}\footnote{%
electronic address: takeuchi@vt.edu}
\\
\medskip
\textit{Institute for Particle Physics and Astrophysics\\
Physics Department, Virginia Tech, Blacksburg, VA 24061--0435}
\\

\bigskip
\bigskip
\bigskip
\bigskip
\bigskip

\begin{abstract}
We use the LEP/SLD data to place constraints on
Topcolor Assisted Technicolor Models. 
We find that due to a large negative shift in
$R_b$ induced by charged top-pion exchange,
it is difficult to make the models 
compatible with experiment.
\end{abstract}

\end{center}

\vfill

\begin{flushleft}
VPI--IPPAP--98--7\\
hep--ph/9812377\\
December 1998
\end{flushleft}

\end{titlepage}

\renewcommand{\thefootnote}{\arabic{footnote}}
\setcounter{footnote}{0}

\begin{flushleft}
\textbf{\large 1. Introduction}
\end{flushleft}
\medskip

In top-color models of electroweak symmetry breaking, the
top-color interaction becomes strong and broken at a scale $\Lambda$.
This generates a top quark condensate
which gives rise to a triplet of Goldstone bosons, 
the top-pions, which are absorbed into the $W^\pm$ and the $Z$.  
In such models, the top-pion decay constant $f_\pi$, which
determines the masses of the $W^\pm$ and the $Z$, and the top mass
$m_t$ are related by \cite{BARDEEN:90}
\begin{equation}
f_\pi^2 = m_t^2 \left( \frac{N_c}{16\pi^2}\ln\frac{\Lambda^2}{\mu^2}
                \right).
\label{EQ1}
\end{equation}
Here, $\mu$ is a scale of the order of $m_t$.  To obtain the
correct masses for the $W^\pm$ and the $Z$, one needs 
$f_\pi = v = 174$~GeV which implies $\Lambda \sim 10^{13-14}$~GeV.
Because of this large hierarchy between $m_t$, $f_\pi$ and $\Lambda$,
top-color models typically require extreme fine tuning of 
the coupling constants to obtain the correct masses for the
gauge bosons and the top.

In Ref.~\cite{HILL:95}, Hill proposed to remedy this
problem by lowering the top-color scale $\Lambda$ to the
order of a TeV.  This lowers the value of $f_\pi$ to
about:
\[    f_\pi \approx 50\,\mathrm{GeV}.   \]
In addition to the top-color interactions, Hill introduced
technicolor \cite{TECHNI}
to generate a condensate of technifermions
with a technipion decay constant $F_\pi$ which satisfies
\[    F_\pi^2 + f_\pi^2 = v^2 = (174\,\mathrm{GeV})^2,      \]
or
\[    F_\pi^2 \approx (167\,\mathrm{GeV})^2.      \]
Thus, the majority of the $W^\pm$ and $Z$ masses come from the
technifermion condensate, while the top quark condensate
serves to make the top quark heavy.   This type of model was
dubbed ``top-color assisted technicolor'' and has been studied 
by many authors \cite{BURDMAN:97,CHIVUKULA:98,POPOVIC:98,
LANE:96,CHIVUKULA:95,KOMINIS:95,HILL:96,SU:97}.

However, it was pointed out by Burdman and Kominis \cite{BURDMAN:97}
that the smallness of the top-pion decay constant $f_\pi$ will
have a dangerous effect on $R_b = \Gamma_{b\bar{b}}/\Gamhad$.
This is because the Yukawa coupling of 
the top quark and the left--handed bottom quark
to the top-pions is given by
\[
y_t = \frac{m_t}{f_\pi} \approx 3.5
\]
which is very large.
Since the top-pions\footnote{Actually, a linear combination of
the top-pions and technipions are absorbed in to the gauge bosons
leaving the linear combination orthogonal to it physical.
The absorbed Goldstone linear combination is mostly the technipion while
the physical linear combination is mostly the top-pion.}
remain unabsorbed and physical in these models,
there is a large radiative correction
to the $Zb\bar{b}$ vertex coming from 
the charged-top-pion -- top-quark loop.

The charged top-pion correction to the $Zb\bar{b}$ vertex
is exactly the same as that of the charged Higgs correction
in two Higgs doublet models with $v_1 = f_\pi$ and $v_2 = F_\pi$.
As discussed by Grant in Ref.~\cite{GRANT:95},
the shift in the left handed coupling of the $b$ to the $Z$ due
to this correction is given by\footnote{We normalize the
coupling so that at tree level, they are given by
\[
g_L^b = -\frac{1}{2} + \frac{1}{3}s^2,\qquad 
g_R^b = \frac{1}{3}s^2.
\]
}
\begin{equation}
\delta g_L^b 
= \frac{1}{2}\left(\frac{y_t^2}{16\pi^2}\right)\frac{v_2^2}{v^2}
  \left[ -\frac{x}{(x-1)^2}\log x + \frac{x}{x-1} \right]
\equiv \tpshift(m_+),
\label{TOPPION-SHIFT}
\end{equation}
where $x = m_t^2/m_+^2$.
The $\frac{1}{2}$ in front is isospin, and the
factor $(v_2^2/v^2)$ is due to top-pion--technipion mixing.
In the limit that the charged Higgs/top-pion
mass $m_+$ goes to infinity, we find that
$\delta g_L^b$ goes to zero, i.e. the contribution decouples.
However, because the Yukawa coupling $y_t$ is so
large, we find that $\delta g_L$ is not small even for
fairly large values of $m_+$.
For instance, if $m_+ = 1$~TeV, we find $\delta g_L = +0.003$.
This amounts to a $+0.7\%$ shift in $g_L$, 
and a $-1.3\%$ shift in $\Gamma_{b\bar{b}}$.
This would shift the theoretical value of
$R_b = \Gamma_{b\bar{b}} / \Gamma_\mathrm{had}$
by $-1\%$ from the Standard Model value of 0.2158
($m_t = 174$~GeV, $m_H = 300$~GeV) down to about 0.2136.
Given that the current experimental value of $R_b$ is \cite{GRUNEWALD:98}
\[
R_b = 0.21656 \pm 0.00074,
\]
the difference would be at the $4\sigma$ level.
For more realistic\footnote{The top-pion is a pseudo--Goldstone boson
whose mass must be generated by ETC \cite{ETC}
interactions.  Hill \cite{HILL:95}
estimates their masses to be around 200~GeV.}
and smaller values of the top-pion mass $m_+$, 
the shift in $g_L^b$, and thus the discrepancy between theory and 
experiment would be huge \cite{BURDMAN:97}.

Since this is a 1--loop calculation
for a Yukawa coupling which is large ($y_t \approx 3.5$),
this result may not be particularly robust.  
However, the 1--loop result does serve as a guideline on
how large the correction can be, and since
the mass of the top-pion $m_+$ can be adjusted,
we can use that freedom to hide our ignorance on the 
higher--order corrections.  
We will therefore refer to the value of
$m_+$ used in Eq.~\ref{TOPPION-SHIFT} as the
\textit{effective} top-pion mass.

Of course, one cannot conclude that
top-color assisted technicolor is ruled out 
on the basis of this observation alone.  
Indeed, it was pointed out by Hill and Zhang \cite{HILL-ZHANG:95} that
coloron dressing of the $Zb\bar{b}$ vertex
actually shifts the left and right handed couplings of the
$b$ to the $Z$ by
\begin{equation}
\frac{\delta g_L^b}{g_L^b} =
\frac{\delta g_R^b}{g_R^b} =
\frac{\kappa_3}{6\pi}C_2(R)
\left[ \frac{m_Z^2}{M_C^2} \ln\frac{M_C^2}{m_Z^2}\right].
\label{COLORON-SHIFT}
\end{equation}
Here, $\kappa_3$ is the coloron coupling (to be defined in the next section)
and $M_C$ is the coloron mass.  
Again, we are using a 1--loop result for a large $\kappa_3$, 
so it should be considered the \textit{effective} coupling
constant for our purpose.
For $M_C \approx 1$~TeV, we find
\[
\frac{\delta g_L^b}{g_L^b} =
\frac{\delta g_R^b}{g_R^b} = 0.003\,\kappa_3.
\]
This leads to a positive shift in $\Gamma_{b\bar{b}}$ and $R_b$ of
\[
\frac{\delta\Gamma_{b\bar{b}}}{\Gamma_{b\bar{b}}} = 0.006\,\kappa_3,\qquad
\frac{\delta R_b}{R_b} = 0.005\,\kappa_3.
\]
If $\kappa_3 \approx 2$, $R_b$ would be shifted to the positive side
by 1\%.  Furthermore, the $Z'$ dressing of the $Zb\bar{b}$ vertex will also
enhance $R_b$,
In principle, therefore, it is possible to 
cancel the large negative top-pion
contribution to $R_b$ with an equally large but positive coloron
and $Z'$ contribution.  
The question is whether such a large correction
is allowed by the other observables or not.

Clearly, one must consider all possible radiative
corrections from all the particles involved and perform a global
fit to the precision electroweak data.
In this paper, we perform a systematic analysis of all
relevant corrections to the $Zf\bar{f}$ vertices.
The $Z$-pole data from LEP and SLD will be used to constrain the size of
these vertex corrections and the effective top-color assisted technicolor
parameters associated with them.

In section~2, we review top-color assisted technicolor 
and introduce the notation.  The version we will be considering
is the one with a strong $U(1)$ interaction ``tilting'' the vacuum.
In section~3, we list all the relevant corrections
we will be considering and discuss how they affect $Z$--pole
observables.
In section~4, we report the result of our fit
to the latest LEP/SLD data.
Section 5 concludes with a discussion on the interpretation of
the result.

\bigskip
\begin{flushleft}
\textbf{\large 2. Top-color Assisted Technicolor}
\end{flushleft}
\medskip

We concentrate our attention to the class of
top-color assisted technicolor models which assume that 
the quarks and leptons transform under the gauge group
\[
SU(3)_s \times SU(3)_w \times 
 U(1)_s \times  U(1)_w \times SU(2)_L
\]
with coupling constants $g_{3s}$, $g_{3w}$, $g_{1s}$, $g_{1w}$,
and $g_2$.  It is assumed that $g_{3s} \gg g_{3w}$ and
$g_{1s} \gg g_{1w}$.\footnote{Several authors have commented that
having a strong $U(1)$ will cause the Landau pole to be
situated not too far from the top-color scale \cite{CHIVUKULA:98,POPOVIC:98}.
We will discuss this problem in a subsequent paper \cite{LOINAZ:99}.}
The charge assignments of the three generation of ordinary fermions
under these gauge groups are given in Table~\ref{CHARGE-ASSIGNMENTS}.
Note that each generation must transform non-trivially under
only one of the $SU(3)$'s and one of the $U(1)$'s, and that those
charges are the same as that of the Standard Model color and hypercharge.
This ensures anomaly cancellation.

\begin{table}[t]
\begin{center}
\begin{tabular}{|c||c|c|c|c|c|}
\hline
           & $SU(3)_s$ & $SU(3)_w$ & $U(1)_s$ & $U(1)_w$ & $SU(2)_L$ \\
\hline\hline
$(t,b)_L$  &  3  &  1  & $\dfrac{1}{3}$  &   0   &   2   \\
\hline
$(t,b)_R$  &  3  &  1  & $\left(\dfrac{4}{3},-\dfrac{2}{3}\right)$
                                        &   0   &   1   \\
\hline
$(\nu_\tau,\tau^-)_L$
           &  1  &  1  & $-1$     &   0   &   2   \\
\hline
$\tau^-_R$ &  1  &  1  & $-2$     &   0   &   1   \\
\hline
$(c,s)_L$, 
$(u,d)_L$  &  1  &  3  &  0  & $\dfrac{1}{3}$   &   2   \\
\hline
$(c,s)_R$, 
$(u,d)_R$  &  1  &  3  &  0  & $\left(\dfrac{4}{3},-\dfrac{2}{3}\right)$
                                                &   1   \\
\hline
$(\nu_\mu,\mu^-)_L$, $(\nu_e,e^-)_L$
           &  1  &  1  &  0  & $-1$  &   2   \\
\hline
$\mu^-_R$, 
$e^-_R$    &  1  &  1  &  0  & $-2$  &   1   \\
\hline
\end{tabular}
\end{center}
\caption{Charge assignments of the ordinary fermions.}
\label{CHARGE-ASSIGNMENTS}
\end{table}

Alternative charge choices to the one shown in
Table~\ref{CHARGE-ASSIGNMENTS} are possible.
In the original model of Hill \cite{HILL:95}, 
the second generation was assigned $U(1)_s$ charges instead
of those under $U(1)_w$
in order to distinguish it from the first generation.
In the model recently proposed by
Popovic and Simmons \cite{POPOVIC:98}, both the first and second
generations were given $SU(3)_s$ quantum numbers instead of 
those under $SU(3)_w$.
Lane \cite{LANE:96} discussed a more general form of
$U(1)_{s,w}$ charge assignments that ensures anomaly cancellation. 
We will comment on the consequences of these alternative
assignments at the end of section~5.

At scale $\Lambda\sim 1$~TeV, technicolor
is assumed to become strong and generate a condensate 
(of something which we will leave unspecified)
with charge $(3,\bar{3},p,-p,1)$ which
breaks the two $SU(3)$'s and the two $U(1)$'s to their diagonal
subgroups 
\[
SU(3)_s \times SU(3)_w \rightarrow SU(3)_c,\qquad
U(1)_s \times U(1)_w \rightarrow U(1)_Y,
\]
which we identify
with the usual Standard Model color and hypercharge groups.

The massless unbroken SU(3) gauge bosons (the gluons $G_\mu^a$) and 
the massive broken SU(3) gauge bosons 
(the so called \textit{colorons} $C_\mu^a$) 
are related to the original $SU(3)_s\times SU(3)_w$ gauge fields
$X_{s\mu}^a$ and $X_{w\mu}^a$ by
\begin{eqnarray*}
C_\mu & = & X_{s\mu} \cos\theta_3 - X_{w\mu} \sin\theta_3 \cr
G_\mu & = & X_{s\mu} \sin\theta_3 + X_{w\mu} \cos\theta_3
\end{eqnarray*}
where we have suppressed the color index, and
\[    \tan\theta_3 = \frac{g_{3w}}{g_{3s}}.     \]
The currents to which the gluons and colorons couple to are:
\[
g_{3s}J_{3s}^\mu X_{s\mu} + g_{3w}J_{3w}^\mu X_{w\mu}
= g_3\left( \cot\theta_3 J_{3s}^\mu - \tan\theta_3 J_{3w}^\mu
     \right) C_\mu
+ g_3\left( J_{3s}^\mu + J_{3w}^\mu \right) G_\mu,
\]
where
\[  \frac{1}{g_3^2} = \frac{1}{g_{3s}^2} + \frac{1}{g_{3w}^2}.  \]
Since the quarks carry only one of the $SU(3)$ charges, we can identify
\[   J_3^\mu = J_{3s}^\mu + J_{3w}^\mu   \]
as the QCD color current, and $g_3$ as the QCD coupling constant.

Similarly, the massless unbroken U(1) gauge boson $B_\mu$ and 
the massive broken U(1) gauge boson $Z'_\mu$
are related to the original $U(1)_s\times U(1)_w$ gauge fields
$Y_{s\mu}$ and $Y_{w\mu}$ by
\begin{eqnarray*}
Z'_\mu & = & Y_{s\mu} \cos\theta_1 - Y_{w\mu} \sin\theta_1 \cr
B_\mu  & = & Y_{s\mu} \sin\theta_1 + Y_{w\mu} \cos\theta_1
\end{eqnarray*}
where
\[    \tan\theta_1 = \frac{g_{1w}}{g_{1s}}.     \]
The currents to which the $B_\mu$ and $Z'_\mu$ couple to are:
\[
g_{1s}J_{1s}^\mu Y_{s\mu} + g_{1w}J_{1w}^\mu Y_{w\mu}
= g_1\left( \cot\theta_1 J_{1s}^\mu - \tan\theta_1 J_{1w}^\mu
     \right) Z'_\mu
+ g_1\left( J_{1s}^\mu + J_{1w}^\mu \right) B_\mu,
\]
where
\[  \frac{1}{g_1^2} = \frac{1}{g_{1s}^2} + \frac{1}{g_{1w}^2}.  \]
Again,
since the fermions carry only one of the $U(1)$ charges, we can identify
\[   J_1^\mu = J_{1s}^\mu + J_{1w}^\mu   \]
as the Standard Model hypercharge current and $g_1$ as the hypercharge
coupling constant.

The masses of the colorons and the $Z'$ will be given by
\begin{eqnarray*}
M_C    & = &    \mathcal{F}\sqrt{g_{3s}^2 + g_{3w}^2},  \cr
M_{Z'} & = & |p|\mathcal{F}\sqrt{g_{1s}^2 + g_{1w}^2}.
\end{eqnarray*}
where $\mathcal{F}$ is the Goldstone boson decay constant associated
with the breaking.
Note that the mass of the $Z'$ can be adjusted at will by
adjusting the charge $p$ of the condensate.

Below the symmetry breaking scale $\Lambda\sim 1$ TeV,
the exchange of the massive colorons and the $Z'$ give rise to
effective four--fermion interactions of the form
\begin{eqnarray*}
\mathcal{L} & = &
- \frac{g_3^2}{2M_C^2}
  \left( \cot\theta_3 J_{3s}^\mu - \tan\theta_3 J_{3w}^\mu \right)
  \left( \cot\theta_3 J_{3s\mu}  - \tan\theta_3 J_{3w\mu}  \right) \cr
&&
- \frac{g_1^2}{2M_{Z'}^2}
  \left( \cot\theta_1 J_{1s}^\mu - \tan\theta_1 J_{1w}^\mu \right)
  \left( \cot\theta_1 J_{1s\mu}  - \tan\theta_1 J_{1w\mu}  \right).
\end{eqnarray*}
Since $\tan\theta_i \ll \cot\theta_i$ ($i=1,3$) by assumption,
we neglect the $J_{iw}$ terms and find
\[
\mathcal{L}
= -\frac{2\pi\kappa_3}{M_C^2}    J_{3s}^\mu J_{3s\mu}
  -\frac{2\pi\kappa_1}{M_{Z'}^2} J_{1s}^\mu J_{1s\mu},
\]
where we have defined
\[   \kappa_i \equiv \frac{g_i^2}{4\pi}\cot^2\theta_i,\qquad (i=1,3).   \]
Note that due to the hypercharge assignments,
the $Z'$ exchange interaction is attractive in the $\bar{t}t$ channel
but repulsive in the $\bar{b}b$ channel while coloron exchange is
attractive in both channels.
Therefore, it is possible to arrange the coupling strengths
$\kappa_3$ and $\kappa_1$ so that the combination of the
coloron and $Z'$ exchange interactions will condense the top, 
but not the bottom.  (This is sometimes called \textit{tilting}
the vacuum.)
Using the Nambu Jona-Lasinio approximation \cite{NJL}, we find that
this requirement places the following constraint on the $\kappa$'s:
\[
C_2(R)\kappa_3 + \frac{1}{9}\kappa_1 > \pi,\qquad
C_2(R)\kappa_3 - \frac{1}{18}\kappa_1 < \pi,
\]
where $C_2(R) = \frac{N_c^2-1}{2N_c}$, $N_c=3$.  
In the large $N_c$ limit, $C_2(R) \approx \frac{N_c}{2} =\frac{3}{2}$,
so the above constraint becomes
\begin{equation}
\kappa_3 + \frac{2}{27}\kappa_1 > \frac{2\pi}{3},\qquad
\kappa_3 - \frac{1}{27}\kappa_1 < \frac{2\pi}{3}.
\label{NJL-LIMIT1}
\end{equation}
In addition, the requirement that the $\tau$ lepton
does not condense leads to
\begin{equation}   
\kappa_1 < 2\pi.
\label{NJL-LIMIT2}   
\end{equation}

Under these conditions, the top quark condensate 
will form $\vev{\bar{t}t}\neq 0$
generating the top quark mass $m_t$ and the top-pions with
decay constant $f_\pi$ which are related through Eq.~\ref{EQ1}.   
This breaks $SU(2)_L\times U(1)_Y$ down
to $U(1)_{em}$, generating (smallish) masses for the $W^\pm$ and the $Z$.
The coupling of the top and bottom quarks to the top-pions is given by
\[
y_t
\left[ \frac{1}{\sqrt{2}}(\bar{t}i\gamma_5 t) \pi_0 + 
       \bar{t}_R b_L \pi^+ + \bar{b}_L t_R \pi^-
\right]
\]
where $y_t = m_t/f_\pi$.

The remainder of the masses of the $W^\pm$ and the $Z$ are assumed
to come from a technifermion condensate in the usual fashion.
The smaller fermion masses are generated through ETC interactions,
including a small ETC mass for the top so that the top-pions will
become massive.

\bigskip
\begin{flushleft}
\textbf{\large 3. Vertex Corrections in Top-color Assisted Technicolor}
\end{flushleft}
\medskip

In previous attempts to constrain top-color assisted technicolor
using precision electroweak measurements \cite{CHIVUKULA:95}
attention had been focussed on the vacuum polarization corrections,
namely the shift in the $\rho$ parameter and $Z$--$Z'$ mixing.

Focussing attention on vacuum polarization corrections
has been the standard technique in analyzing precision electroweak
data \cite{PESKIN:90}.
The main advantage in doing this is that vacuum polarization
corrections modify the gauge boson propagators and are therefore
universal: they correct \textit{all} electroweak observables
and therefore \textit{all} the electroweak data can be used to
constrain their sizes.

However, there are serious disadvantages also.
First, each gauge boson couples to all particles that carry
its charge so that the model under consideration must be
specified \textit{completely}.   
In top-color assisted technicolor models, this means that the charges and
masses of the techni-sector must be specified
which makes any limit highly model dependent.
Second, in order to be able to use all electroweak data
to constrain the vacuum polarization corrections, one often neglects
the highly process--dependent vertex and box corrections 
which may not be negligible at all.
In Ref.~\cite{CHIVUKULA:95}, the only corrections considered
were vacuum polarization corrections coming from technifermions
of specific models.   Vacuum polarization and vertex corrections
coming from ordinary fermion and top-pion loops were completely
neglected.

A much better way to deal with top-color assisted technicolor
and similar theories
is to focus on \textit{vertex corrections at the $Z$ pole only}.
This allows us to place severe constraints on the theory
without specifying the technisector.  All that is necessary is
to specify the charges of the ordinary quarks and leptons.
(A similar technique was used in Ref.~\cite{DPF:94} to
constrain corrections to the $Zb\bar{b}$ vertex.)

Let us now list the vertex corrections that must be considered.
They come in two classes, namely:
\begin{enumerate}
\item gauge boson mixing terms, and
\item proper vertex corrections.
\end{enumerate}
Gauge boson mixing corrections to the $Zf\bar{f}$ vertices are due
to the rediagonalization of the gauge bosons from vacuum polarization
corrections.
At tree level, the $Z$ couples to the current
\[
J_Z^0 = J_{I_3} - s^2 J_Q,
\]
where $s^2$ is shorthand for $\sin^2\theta_w$.
$Z$--photon mixing and $Z$--$Z'$ mixing will modify this
current to:
\[
J_Z = J_{I_3} - (s^2 + \ds) J_Q + \epsilon J_{1s},
\]
where $\ds$ and $\epsilon$ parametrize the size of
the $Z$--photon and $Z$--$Z'$ mixings, respectively.
We neglect the small $J_{1w}$ component of the $J_{Z'}$
current.  We need not worry about the overall change in scale
due to these corrections since the observables
we will be looking at are all \textit{ratios of coupling constants}
from which such scale dependence vanishes.

Since we will be using only $Z$--pole observables in our analysis,
$\ds$ and $\epsilon$ will remain phenomenological
parameters and will
not yield any information on the vacuum polarization
corrections which give rise to them.   Vacuum polarizations are
visible only when comparing processes at different energy scales,
or processes involving different gauge bosons.\footnote{
For instance, the $S$ parameter is only visible when comparing
neutral current processes at different energy scales and the $T$ parameter
is only visible when comparing neutral and charged current processes.}

The proper vertex corrections we must consider are
the top-pion and coloron corrections discussed in the
introduction and the $Z'$ dressing corrections.
We neglect all other corrections that vanish in the limit that
all the fermion masses (except that of the top) are
taken to zero.
We also make the simplifying assumption that
the bottom-pions \cite{KOMINIS:95} are heavy enough so that 
their effects are negligible.

Since the couplings of the colorons and the $Z'$ to the
$SU(3)_w$ and $U(1)_w$ charges are highly suppressed,
they can also be neglected.  
Then, with the charge assignment given in Table~\ref{CHARGE-ASSIGNMENTS},
the only vertices that receive coloron and
$Z'$ dressing corrections are $Zb\bar{b}$ and $Z\tau^+\tau^-$. 
The coloron correction was given in Eq.~\ref{COLORON-SHIFT}, and
the $Z'$ correction can be obtained by simply replacing
$\kappa_3$ and $M_C$ with $\kappa_1$ and $M_{Z'}$, respectively,
and the color factor $C_2(R) = \frac{N^2-1}{2N} = \frac{4}{3}$ 
by the hypercharge squared:
\begin{eqnarray*}
\frac{\delta g_L(f)}{g_L(f)}
& = & \frac{\kappa_1}{6\pi}(Y_L^f)^2
      \left[ \frac{m_Z^2}{M_{Z'}^2} \ln\frac{M_{Z'}^2}{m_Z^2} \right],\cr
\frac{\delta g_R(f)}{g_R(f)}
& = & \frac{\kappa_1}{6\pi}(Y_R^f)^2
      \left[ \frac{m_Z^2}{M_{Z'}^2} \ln\frac{M_{Z'}^2}{m_Z^2} \right].
\end{eqnarray*}
In the following, we will use $M_C = M_{Z'} = 1$~TeV.

Therefore, the couplings of
the first and second generation fermions only receive 
corrections from photon-$Z$ mixing:
\begin{eqnarray*}
\dgl(\nu_e) \;=\; \dgl(\nu_\mu) & = & 0 \cr
\dgl(e) \;=\; \dgl(\mu) \;=\; \dgr(e) \;=\; \dgr(\mu) & = & \ds  \cr
\dgl(u) \;=\; \dgl(c) \;=\; \dgr(u) \;=\; \dgr(c) & = & -\frac{2}{3}\ds \cr
\dgl(d) \;=\; \dgl(s) \;=\; \dgr(d) \;=\; \dgr(s) & = & \phantom{-}\frac{1}{3}\ds \cr
\end{eqnarray*}
while the couplings of the third generation fermions receive
all corrections:
\begin{eqnarray*}
\dgl(\nu_\tau)
& = &     -  \epsilon + 0.0021 \kappa_1 g_L(\nu_\tau) \cr
\dgl(\tau)
& = & \ds -  \epsilon + 0.0021 \kappa_1 g_L(\tau) \cr
\dgr(\tau)
& = & \ds - 2\epsilon + 0.0085 \kappa_1 g_R(\tau) \cr
\dgl(b)
& = & \frac{1}{3}\ds + \frac{1}{3}\epsilon 
    + (0.00023\kappa_1 + 0.0028\kappa_3)g_L(b) + \tpshift(m_+) \cr
\dgr(b)
& = & \frac{1}{3}\ds - \frac{2}{3}\epsilon 
    + (0.00094\kappa_1 + 0.0028\kappa_3)g_R(b)
\end{eqnarray*}
Here, $\tpshift(m_+)$ denotes the top-pion correction.

Given these expressions, we can now
calculate how the $Z$--pole observables are shifted
by non-zero values of $\ds$, $\epsilon$, $\kappa_1$ and $\kappa_3$,
and fit the result to the experimental data.
We will also let the QCD coupling constant $\alpha_s(m_Z)$ float
in our fit so that the size of the QCD gluon dressing corrections can
be adjusted.  We define the parameter 
$\da$ to be the shift of $\alpha_s(m_Z)$ away from
its nominal value of 0.120:
\[         \alpha_z(m_Z) = 0.120 + \da.      \]

\newpage

\begin{table}[ht]
\begin{center}
\begin{tabular}{|c|c|c|}
\hline
Observable & Measured Value & ZFITTER Prediction \\
\hline\hline
\multicolumn{2}{|l|}{\underline{$Z$ lineshape variables}} & \\
$m_Z$        & $91.1867 \pm 0.0021$ GeV & input       \\
$\GamZ$      & $2.4939 \pm 0.0024$ GeV  & unused      \\
$\sighad$    & $41.491 \pm 0.058$ nb    & $41.468$ nb \\
$R_e$        & $20.783 \pm 0.052$       & $20.749$ \\
$R_\mu$      & $20.789 \pm 0.034$       & $20.749$ \\
$R_\tau$     & $20.764 \pm 0.045$       & $20.796$ \\
$\AFB(e)$    & $0.0153 \pm 0.0025$      & $0.0154$ \\
$\AFB(\mu)$  & $0.0164 \pm 0.0013$      & $0.0154$ \\
$\AFB(\tau)$ & $0.0183 \pm 0.0017$      & $0.0154$ \\
\hline
\multicolumn{2}{|l|}{\underline{$\tau$ polarization}} & \\
$A_e$        & $0.1479 \pm 0.0051$      & $0.1433$ \\ 
$A_\tau$     & $0.1431 \pm 0.0045$      & $0.1435$ \\
\hline
\multicolumn{2}{|l|}{\underline{SLD left--right asymmetry}} & \\
$A_e$        & $0.1504 \pm 0.0023$      & $0.1433$ \\
\hline
\multicolumn{2}{|l|}{\underline{heavy flavor observables}} & \\
$R_b$        & $0.21656 \pm 0.00074$    & $0.2158$ \\
$R_c$        & $0.1735  \pm 0.0044$     & $0.1723$ \\
$\AFB(b)$    & $0.0990  \pm 0.0021$     & $0.1004$ \\
$\AFB(c)$    & $0.0709  \pm 0.0044$     & $0.0716$ \\
$A_b$        & $0.867   \pm 0.035$      & $0.934$ \\
$A_c$        & $0.647   \pm 0.040$      & $0.666$ \\ 
\hline
\end{tabular}
\caption{LEP/SLD observables \cite{GRUNEWALD:98} 
and their Standard Model predictions.
The predictions were calculated using ZFITTER \cite{ZFITTER:92} 
with $m_t = 173.9$~GeV,
$m_H = 300$~GeV, $\alpha_s(m_Z) = 0.120$, and $\alpha^{-1}(m_Z) = 128.9$.}
\label{LEP-SLD-DATA}
\end{center}
\end{table}

\bigskip
\begin{flushleft}
\textbf{\large 4. Fit to LEP/SLD Data}
\end{flushleft}
\medskip

In Table~\ref{LEP-SLD-DATA} we show the latest LEP/SLD data 
obtained from Ref.~\cite{GRUNEWALD:98}.  The correlation matrices for
the errors in the $Z$-lineshape variables and the heavy flavor
observables are shown in the appendix.

Of the 9 lineshape variables, the three $R_\ell$ ratios and
the three forward-backward asymmetries are just ratios of coupling constants.
Of the remaining three, the product
\[
m_Z^2\sighad = 12\pi \frac{\Game\Gamhad}{\GamZ^2}
\]
is again just a ratio of coupling constants.
All the other observables shown in Table~\ref{LEP-SLD-DATA} are
ratios of coupling constants.

We therefore have 16 observables which we can use in our analysis.
The shifts in these observables due to $\tpshift(m_+)$
and non-zero values of $\ds$, $\epsilon$, $\kappa_1$, $\kappa_3$, 
and $\da$ are:
\begin{eqnarray}
\frac{\delta\sighad}{\sighad}
& = & 0.11\,\ds + 0.93\,\epsilon
      - 0.0013\,\kappa_1
      - 0.0005\,\kappa_3
      - 0.12\,\da + 0.40\,\tpshift                     \cr
\frac{\delta R_e}{R_e} \;=\;\frac{\delta R_\mu}{R_\mu}
& = & - 0.86\,\ds - 0.46\,\epsilon 
      + 0.0001\,\kappa_1
      + 0.0012\,\kappa_3                      
      + 0.31\,\da - 1.0\,\tpshift                      \cr
\frac{\delta R_\tau}{R_\tau}
& = & - 0.86\,\ds + 2.7\,\epsilon
      - 0.0096\,\kappa_1
      + 0.0012\,\kappa_3                     
      + 0.31\,\da - 1.0\,\tpshift                      \cr
\frac{\delta\AFB(e)}{\AFB(e)} &=& \frac{\delta\AFB(\mu)}{\AFB(\mu)}
\;=\; - 110\,\ds                                             \cr
\frac{\delta\AFB(\tau)}{\AFB(\tau)}
& = & - 110\,\ds + 84\,\epsilon - 0.043\,\kappa_1             \cr
\frac{\delta A_e}{A_e}
& = &  - 55\,\ds                                             \cr
\frac{\delta A_\tau}{A_\tau}
& = &  - 55\,\ds + 84\,\epsilon - 0.043\,\kappa_1             \cr
\frac{\delta R_b}{R_b}
& = &  0.18\,\ds - 1.6\,\epsilon 
     + 0.0004\,\kappa_1
     + 0.0044\,\kappa_3 - 3.6\,\tpshift  \cr
\frac{\delta R_c}{R_c}
& = & -0.35\,\ds + 0.46\,\epsilon
     - 0.0001\,\kappa_1
     - 0.0012\,\kappa_3 + 1.0\,\tpshift  \cr
\frac{\delta\AFB(b)}{\AFB(b)}
& = & -56\,\ds + 1.1\,\epsilon
     - 0.00009\,\kappa_1
     - 0.32\,\tpshift                                 \cr
\frac{\delta\AFB(c)}{\AFB(c)}
& = & -60\,\ds                                         \cr
\frac{\delta A_b}{A_b}
& = & -0.68\,\ds +1.1\,\epsilon
     - 0.00009\,\kappa_1
     - 0.32\,\tpshift                                 \cr
\frac{\delta A_c}{A_c}
& = & -5.2\,\ds
\label{FIT-COEFS}
\end{eqnarray}
The top-pion correction $\tpshift(m_+)$ in these expressions
is fixed by choosing an effective top-pion mass $m_+$.
The remaining 5 parameters: $\ds$, $\epsilon$, $\kappa_1$,
$\kappa_3$, and $\da$ are
fit to the data given in Table~\ref{LEP-SLD-DATA},
taking into account the correlations between the experimental
errors given in the appendix.

\begin{figure}[p]
\centering
\unitlength=1cm
\begin{picture}(13,10.5)
\unitlength=1mm
\put(90,90){$m_+ = 1000$ GeV}
\put(79,70){$R_b$}
\put(110,33){$R_\tau$}
\epsfbox[40 60 412 355]{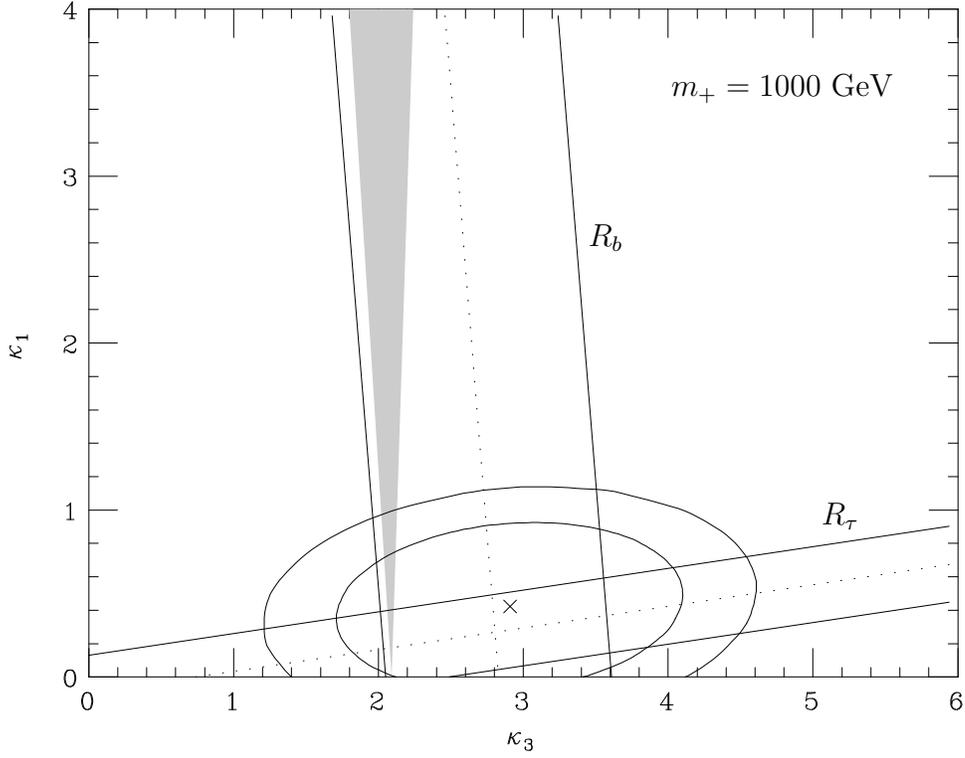}
\end{picture}

\begin{picture}(13,10.5)
\unitlength=1mm
\put(18,90){$m_+ = 600$ GeV}
\put(87,70){$R_b$}
\put(60,20){$R_\tau$}
\epsfbox[40 60 412 355]{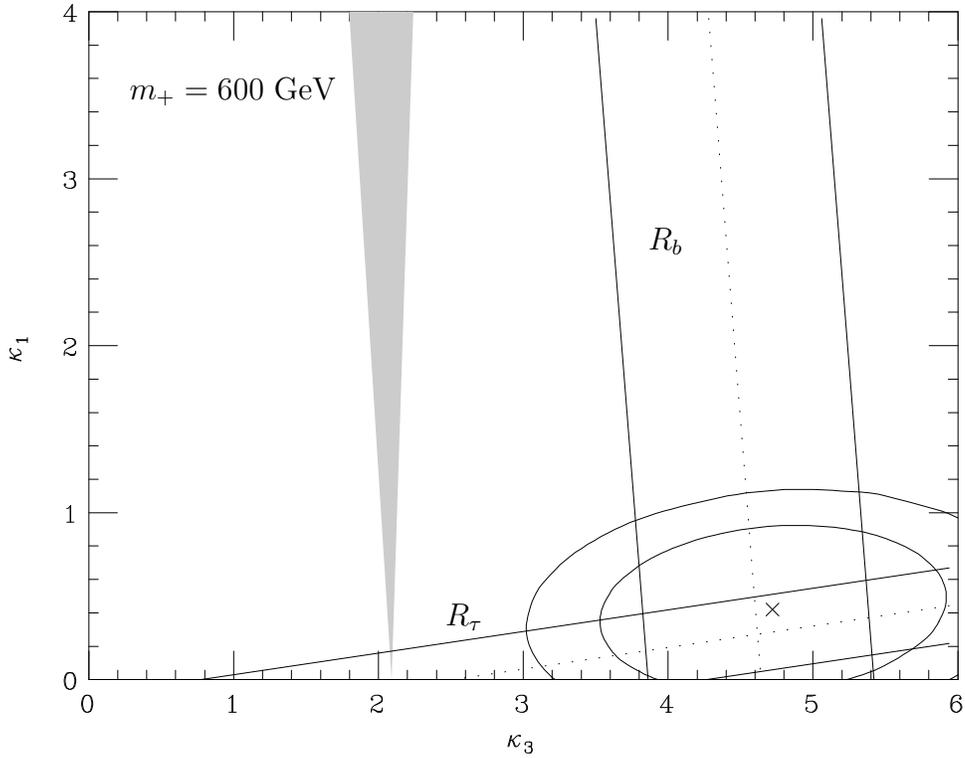}
\end{picture}
\caption{Limits on $\kappa_1$ and $\kappa_3$ for the
$m_+ = 1000$~GeV and $m_+ = 600$~GeV cases.  
The coutours show the 68\% and 90\% confidence limits.  
The shaded area is the region allowed
by the requirement of vacuum tilting.}
\label{FIG1000}
\end{figure}

We choose two reasonable values for the effective top-pion mass: 
$m_+ = 600$ and $1000$~GeV.   
The value of $\tpshift$ for these masses are
\begin{eqnarray*}
\tpshift(\phantom{0}600\,\mathrm{GeV}) & = & 0.006 \cr
\tpshift(1000\,\mathrm{GeV}) & = & 0.003
\end{eqnarray*}
The result of the fit for the $m_+ = 1000$~GeV case is:
\begin{eqnarray*}
\ds       & = & -0.0004 \pm 0.0002 \cr
\epsilon  & = &  0.0005 \pm 0.0005 \cr
\kappa_1  & = &  0.43   \pm 0.33   \cr
\kappa_3  & = &  2.9    \pm 0.8    \cr
\da       & = & -0.0008 \pm 0.0050
\end{eqnarray*}
with the correlation matrix shown in Table~\ref{CORRELATION}.
The quality of the fit was $\chi^2 = 12.6/(16-5)$.
The strongest constraint on $\kappa_3$ comes from $R_b$,
and the strongest constraint on $\kappa_1$ comes from $R_\tau$.
This is shown in Fig.~\ref{FIG1000}.

\begin{table}[ht]
\begin{center}
\begin{tabular}{|c|ccccc|}
\hline
& $\ds$  & $\epsilon$ & $\kappa_1$ & $\kappa_3$ & $\da$ \\
\hline\hline
$\ds$      & $1.00$ & $0.34$ & $0.19$ & $0.09$ & $\phantom{-}0.13$ \\
$\epsilon$ &        & $1.00$ & $0.74$ & $0.23$ & $\phantom{-}0.17$ \\
$\kappa_1$ &        &        & $1.00$ & $0.15$ & $\phantom{-}0.29$ \\
$\kappa_3$ &        &        &        & $1.00$ & $-0.57$           \\
$\da$      &        &        &        &        & $\phantom{-}1.00$ \\
\hline
\end{tabular}
\caption{The correlation matrix of the fit parameters.}
\label{CORRELATION}
\end{center}
\end{table}

As is evident from the figure, the region allowed by our fit overlaps
with the region allowed by the vacuum tilting constraint:
Eqs.~\ref{NJL-LIMIT1} and \ref{NJL-LIMIT2}.
This is as expected from our discussion in the introduction:
if the top-pion mass is large enough, then the top-pion correction
is small enough to be cancelled by the coloron correction.
Though the $Z'$ correction can also be used to cancel the 
top-pion correction in $R_b$, it is suppressed by lepton universality.

For the $m_+ = 600$~GeV case, the fit result is identical to
the $m_+ = 1000$~GeV case except for the limit on $\kappa_3$ which is
\[
\kappa_3 = 4.7 \pm 0.8
\]
This is also shown in Fig.~\ref{FIG1000}.
Obviously, to cancel the top-pion correction, one must move
out of the region allowed by the vacuum tilting constraint.

\bigskip
\begin{flushleft}
\textbf{\large 5. Discussion and Conclusion}
\end{flushleft}
\medskip

Our result demonstrates that the class of
top-color assisted technicolor models we have considered
is ruled out unless:
\begin{enumerate}
\item the effective top-pion mass is around a TeV.
This means that either the higher--order corrections must
suppress the 1--loop correction significantly, or that the
top-pion is indeed as heavy as a TeV, or
\item 1--loop coloron corrections are enhanced significantly
by higher--order corrections.
\end{enumerate}

Hill \cite{HILL:96} suggests that the top-pion
contribution may be sufficiently suppressed by 
taking the top-pion decay constant $f_\pi\approx 100$~GeV.  
This will decrease the Yukawa coupling by a factor of 2
and suppress the top-pion correction by a factor of 4.
However, this requires the top-color scale to be about
$\Lambda\sim 1000$~TeV.  This increase in the top-color
scale will suppress enormously the coloron and $Z'$ corrections,
depriving them of any power to counteract the top-pion correction.
Furthermore, an increase in top-color scale implies the
necessity of fine tuning which is contrary to the original
motivation of the theory. 

It is interesting to note that the experimental values of
$\AFB(b)$ and $A_b$ actually \textit{prefer} a large top-pion correction.
In fact, these two observables contribute the most ($6.8$) to the overall
$\chi^2$ of the fit because the top-pion correction is not
large enough to make the agreement better.\footnote{
The bottom-pion correction, which we have neglected, may
account for the deviation in $\AFB(b)$ and $A_b$.}
Therefore, finding a way to enhance the coloron correction
may be the more phenomenologically viable path.

Since we have examined a model with a specific charge assignment,
one can ask whether a different charge assignment may
improve the situation.  We have looked at several alternative scenarios
and have found the following:
\begin{enumerate}

\item
In the original formulation by Hill \cite{HILL:95},
the second generation was assigned $U(1)_s$ charges instead of
$U(1)_w$ charges.   This assignment would make $\kappa_1$ break
lepton universality between the electron and the muon.
As a result, the limits on $\kappa_1$ will be even tighter
than when only the third generation carried the $U(1)_s$ charge.

\item
In the model recently proposed by Popovic and Simmons \cite{POPOVIC:98},
all three generations were assigned $SU(3)_s$ charges.
This makes the coloron correction cancel exactly in the ratio
$R_b = \Gamma_{b\bar{b}}/\Gamhad$ so it cannot counteract the
top-pion correction at all.

\item
One can free $\kappa_1$ from the constraint of lepton universality
if all three generations are assigned equal $U(1)_s$ charges
and no $U(1)_w$ charge.
However, that would make the $Z'$ correction \textit{decrease} the
ratio $R_b = \Gamma_{b\bar{b}}/\Gamhad$ since the
denominator will grow faster than the numerator.
\footnote{One should also take into account the effect of direct
$Z'$ exchange \cite{SU:97} between the initial $e^+e^-$ pair
and the final $f\bar{f}$ pair in such models, 
but this was not done here.}
\end{enumerate}

There are of course other charge assignments that
one can think of as was considered by Lane \cite{LANE:96}.
However, we feel that these examples more than aptly 
show that changing the charge assignments probably 
will not alleviate the problem.

To summarize:
we have used the latest LEP/SLD data to
place constraints on the size of relevant vertex
corrections to $Z$--pole observables in
top-color assisted technicolor models with a strong
vacuum tilting $U(1)$.
We find that
it is difficult to make the models compatible with experiment
unless the large top-pion correction to $R_b$ can be 
suppressed, or the coloron correction enhanced.

\bigskip
\begin{flushleft}
\textbf{\large Acknowledgements}
\end{flushleft}
\medskip

We would like to thank Martin~W.~Gr\"unewald for supplying us with 
the LEP/SLD data used in this analysis, 
and Aaron~K.~Grant for helpful discussions.
This work was supported in part (W.L.) by the 
U.~S. Department of Energy, grant DE-FG05-92-ER40709, 
Task A.

\bigskip
\bigskip
\bigskip

\newpage
\appendix

\begin{flushleft}
\textbf{\large Appendix: Correlations of LEP/SLD Data}
\end{flushleft}
\medskip

\begin{table}[ht]
\begin{center}
\begin{tabular}{|c|ccccccccc|}
\hline
& $m_Z$     & $\GamZ$     & $\sighad$
& $R_e$     & $R_\mu$     & $R_\tau$ 
& $\AFB(e)$ & $\AFB(\mu)$ & $\AFB(\tau)$ \\
\hline
$m_Z$ 
& $1.000$   & $0.000$     & $-0.040$ 
& $\phantom{-}0.002$ & $-0.010$ & $-0.006$ & $\phantom{-}0.016$ 
& $\phantom{-}0.045$ & $\phantom{-}0.038$ \\
$\GamZ$
& & $\phantom{-}1.000$ & $-0.184$ & $-0.007$ & $\phantom{-}0.003$ 
& $\phantom{-}0.003$ & $\phantom{-}0.009$ & $\phantom{-}0.000$ 
& $\phantom{-}0.003$ \\
$\sighad$
& & & $\phantom{-}1.000$ & $\phantom{-}0.058$ & $\phantom{-}0.094$ 
& $\phantom{-}0.070$ & $\phantom{-}0.006$ & $\phantom{-}0.002$ 
& $\phantom{-}0.005$ \\
$R_e$ & & & & $\phantom{-}1.000$ & $\phantom{-}0.098$ & $\phantom{-}0.073$ 
& $-0.442$ & $\phantom{-}0.007$ & $\phantom{-}0.012$ \\
$R_\mu$ & & & & & $\phantom{-}1.000$ & $\phantom{-}0.105$ 
& $\phantom{-}0.001$ & $\phantom{-}0.010$ & $-0.001$ \\
$R_\tau$ & & & & & & $\phantom{-}1.000$ & $\phantom{-}0.002$ 
& $\phantom{-}0.000$ & $\phantom{-}0.020$ \\
$\AFB(e)$ & & & & & & & $\phantom{-}1.000$ & $-0.008$ & $-0.006$ \\
$\AFB(\mu)$ & & & & & & & & $\phantom{-}1.000$ & $\phantom{-}0.029$ \\
$\AFB(\tau)$ & & & & & & & & & $\phantom{-}1.000$ \\
\hline
\end{tabular}
\caption{The correlation of the $Z$ lineshape variables at LEP}
\end{center}
\end{table}

\begin{table}[h]
\begin{center}
\begin{tabular}{|c|cccccc|}
\hline
& $R_b$ & $R_c$ & $\AFB(b)$ & $\AFB(c)$ & $A_b$ & $A_c$ \\
\hline
$R_b$ & $1.00$ & $-0.17$ & $-0.06$ & $\phantom{-}0.02$ & $-0.02$ 
& $\phantom{-}0.02$ \\
$R_c$ & & $\phantom{-}1.00$ & $\phantom{-}0.05$ & $-0.04$ 
& $\phantom{-}0.01$ & $-0.04$ \\
$\AFB(b)$ & & & $\phantom{-}1.00$ & $\phantom{-}0.13$ & $\phantom{-}0.03$ 
& $\phantom{-}0.02$ \\
$\AFB(c)$ & & & & $\phantom{-}1.00$ & $-0.01$ & $\phantom{-}0.07$ \\
$A_b$     & & & & & $\phantom{-}1.00$ & $\phantom{-}0.04$ \\
$A_c$     & & & & & & $\phantom{-}1.00$ \\
\hline
\end{tabular}
\caption{The correlation of the heavy flavor observables at LEP/SLD.}
\end{center}
\end{table}


\end{document}